\documentclass[aps,prd,onecolumn,showpacs,nofootinbib,amsmath,amssymb,amsfonts,showkeys]{revtex4}
\usepackage{graphicx}
\usepackage{bm}
\usepackage{xcolor}

\begin{document}
\title{Dynamics of dark energy \\ in the gravitational fields of matter inhomogeneities}

\author{Bohdan Novosyadlyj}
 \email{novos@astro.franko.lviv.ua}
\author{Yurij Kulinich}
 \email{kul@astro.franko.lviv.ua}
 \author{Maksym Tsizh}
 \email{tsizh@astro.franko.lviv.ua}
\affiliation{Astronomical Observatory of
Ivan Franko National University of Lviv, \\ Kyryla i Methodia str., 8, Lviv, 79005, Ukraine}
\date{\today}
\begin{abstract}
 We study the dynamical properties and space distribution of dark energy in the weak and strong gravitational fields caused by inhomogeneities of matter in the static world of galaxies and clusters. We show that the dark energy in the weak gravitational fields of matter density perturbations can condense or dilute, but amplitudes of its perturbations remain very small on all scales. We illustrate also how the ``accretion'' of the phantom dark energy onto the matter overdensity forms the dark energy underdensity.
 We analyze the behavior of dark energy in the gravitational fields of stars and black holes with the Schwarzschild metric. It is shown that, in the case of stars, the static solution of the differential equations for energy-momentum conservation exists and describes the distribution of density of dark energy inside and outside of a star. We have found that for stars and galaxies its value differs slightly from the average and is a bit higher for the quintessential scalar field as dark energy and a bit lower for the phantom one. The difference grows with the decrease of the effective sound speed of dark energy and is large in the neighborhood of neutron stars.
We obtain and analyze also the solutions of equations that describe the stationary accretion of the dark energy as a test component onto the Schwarzschild black hole. It is shown that the rate of change of mass of the dark energy is positive in the case of quintessential dark energy and is negative in the case of the phantom one.
\end{abstract}
\maketitle

\section{Introduction}
Astrophysicists have enough evidence that our Universe is expanding with positive acceleration, instead of deceleration, as it should be in a world filled only with ordinary matter ruled by usual laws of gravity. The reality of this fact is beyond doubt, although its interpretation is far from having a single meaning (see, for example, books and review articles \cite{GRG2008,Amendola2010,Wolchin2010,Ruiz2010,Novosyadlyj2013m} devoted to this problem).
One of the most developed interpretations in the terms of theoretical modeling, comparing of numerical predictions with observational data, and determination of the parameters and their confidence intervals is a scalar field that fills the Universe almost homogeneously and slowly rolls down to the minimum of its own potential in the case of a quintessential scalar field or slowly rolls up to the maximum in the case of a phantom one.
There are many possible realizations of such a field (one can find a large portion of references in the sources mentioned above), which is why the additional observational or experimental tests, which could help to constrain the number of candidates at least in the class of scalar field models of dark energy, are needed.
It seems that observational cosmological data obtained in the last few years \cite{wmap9a,Planck2013b,Percival2010,Anderson2012,Padmanabhan2012,wigglez,6dF,snls3,union,Riess2009,Steigman2007,Wright2007} prefer phantom dark energy \cite{Novosyadlyj2012,Novosyadlyj2013,Xia2013,Cheng2013,Rest2013,Shafer2013,Novosyadlyj2014,Hu2014}.
However, we are still far from establishing the physical nature of dark energy, and it is necessary to search for new sensitive tests for it.

The properties of dark energy are extensively investigated in the FRW space-time as the cosmological background and a lot of models have been proposed. The physical parameters of dynamical dark energy (energy density, potential etc.) are time dependent in such a space-time. And contrarily, the properties of dynamical dark energy in the static world of gravitationally bound systems are unknown. We try to remove this lack by analysis
of dynamical properties of the scalar field as the dark energy of either quintessential or phantom type in the galaxies and vicinity of stars and black holes.

The goal of this paper is to study the distribution of the density of quintessential and phantom dark energy in the weak and strong gravitational fields of the static astrophysical objects. In this regard we are interested Refs. \cite{Babichev2004}, the authors of which drew the conclusion about decreasing the mass of a black hole onto which the phantom dark energy falls. To verify this conclusion, we analyze the evolution of density perturbations of dark energy in the weak and strong gravitational fields, the possibility of static configurations of the fields in neighborhood of massive astrophysical objects, and the accretion of the dark energy onto them. We prove the illusiveness of decreasing the black hole mass caused by accretion of the phantom dark energy.

\section{Evolution of dark energy perturbations in the field of dark matter ones}

The numerous studies of the evolution of dark energy perturbations in the expanding Universe, summarized in many reviews and books (see, for example, Refs. \cite{GRG2008,Amendola2010,Wolchin2010,Ruiz2010,Novosyadlyj2013m}), support that - 1) dynamical dark energy with equation of state (EoS) parameter $w\ne-1$ ($p_{de}=w\rho_{de}$) is perturbed due to its own gravitational instability and perturbations of other components - and 2) the amplitudes of density perturbations of dark energy depend on effective sound speed and in most of the viable models are much less than the amplitudes of the density perturbations of matter.  In the paper \cite{Novosyadlyj2014a} we have analyzed the behavior of density and velocity perturbations of dark energy in a static world with the Minkowski metric and have shown that density and velocity perturbations of dark energy  are extremely small in comparison to the dark  matter perturbations on all scales interesting for astrophysics and cosmology. That is, in the weak
gravitational fields the amplitudes of perturbations of dark energy are small and equations for their evolution can be linearized with respect to the perturbed quantities of dark energy. In the galaxy or cluster of galaxies region, where the gravitational field caused by inhomogeneities of matter is weak, the space-time metric can be represented as
\begin{equation}
ds^2=(1+2\Psi(x^i))d\tau^2-(1-2\Psi(x^i))\delta_{\alpha\beta}dx^{\alpha}dx^{\beta},\label{ds}
\end{equation}
where $\Psi(x^i)\ll 1$ is the gravitational potential caused by density inhomogeneities of matter and dark energy (we consider only the scalar mode of perturbations). Here and below, we assume the speed of light to be equal to 1 ($c=1$). The linear theory of perturbations gives exact solutions for the Fourier amplitudes of density and velocity perturbations of dark energy in the field of matter perturbations \cite{Novosyadlyj2014a}, which for scales of structures in galaxies and clusters of galaxies are:
\begin{eqnarray}
\delta_{de}&=&\tilde{C}_1\sin{(c_sk\tau)}+\tilde{C}_2\cos{(c_sk\tau)}-\frac{(1+w)}{1+c_s^2k^2\tau_m^2}\tau_m^2k^2\Psi_k,\label{delta_de}\\
V_{de}&=& -\frac{c_s}{1+w}\left[\tilde{C}_1\cos{(c_sk\tau)}-\tilde{C}_2\sin{(c_sk\tau)}\right]+\frac{(1+3c_s^2)}{1+c_s^2k^2\tau_m^2}\tau_mk\Psi_k, \label{V_de}
\end{eqnarray}
where $\tilde{C}_1$, $\tilde{C}_2$ are constants of integration, which are set by initial conditions, $c_s$ is the effective sound speed (speed of propagation of perturbations) in units of the speed of light, $\tau_m = (4\pi G\rho_m)^{-1/2}$ is the characteristic time scale of matter density perturbations, and $\Psi_k=-\tau_m^{-2}k^{-2}\delta_m(k)$ is the Fourier amplitude of gravitational potential related to the Fourier amplitude of the matter density perturbation $\delta_m(k)\equiv\mathcal{F}\{(\rho_m(r)-\overline{\rho}_m)/\overline{\rho}_m\}$. For the average densities of matter\footnote{They have the dimension of the energy density in formulas.} $\overline{\rho}_{m}$ in galaxies  $\approx 5\cdot10^{-24}$ g/cm$^3$ and in clusters of galaxies  $\approx 5\cdot10^{-27}$ g/cm$^3$, estimated in the framework of the halo model of galaxy formation (see Refs. \cite{Kulinich2013} and citing therein), $\tau_m^{gal}\approx 4.7\, \textrm{Mpc}$ and $\tau_m^{cl}\approx 150\, \textrm{Mpc}$ respectively. Estimations of 
the averaged density of
dark
energy in galaxies and clusters of galaxies give  $\overline{\rho}_{de}^{(gal,\,cl)}\ll \overline{\rho}_{m}^{(gal,\,cl)}$ \cite{Novosyadlyj2014a}. For galaxy or cluster elements of structure, $\tau_mk\gg1$, where $k^{-1}$ is the characteristic scale of the element the of structure. It means that for dark energy models with $c_s\sim1$ another important relation $\delta_{de}\ll\delta_m$ is valid at all scales interesting for astrophysics. The oscillation modes of the solutions (\ref{delta_de}), (\ref{V_de}) reflect the gravitational stability of dark energy beyond the matter inhomogeneity, the gravitational field of which plays the role of driving force in the equations of evolution of dark energy perturbations (see Ref. \cite{Novosyadlyj2014a} for details).

For analysis of the behavior of dark energy in the weak gravitational fields, let us neglect the oscillatory component accepting $\tilde{C}_1=\tilde{C}_2=0$ and analyze only the ``forced'' component of the solutions.  It is determined by the magnitude of the gravitational potential, by its time dependence and also by the parameters of dark energy, EoS parameter $w$, and effective sound speed $c_s$.
For positive perturbations of matter, overdensities of continuous media or separate objects we have always $\Psi_k<0$. The amplitude of forced density perturbations of quintessential dark energy ($w>-1$) has positive sign, and for the phantom one ($w<-1$), it has negative sign. The amplitude of velocities of both types of dark energy in the region of matter overdensities has a negative sign, this means that dark energy flows into a region of matter perturbation, the amplitude of which grows. In other words, during inflow (accretion) of phantom dark energy its density decreases. The physical nature of such decreasing is the same as in the case of the increasing of density of phantom dark energy in the expanding Universe.
 
Although the amplitudes of matter overdensities in the static world of galaxies increase exponentially with time (see solutions (16)-(18) in Ref. \cite{Novosyadlyj2014a}),
$$\delta_m=-\tau_m^2k^2A_k e^{\tau/\tau_m}>0, \quad \Psi_k(\tau)=A_k e^{\tau/\tau_m}<0,$$
the inflow of phantom dark energy according to (\ref{delta_de}) and (\ref{V_de}) forms the dark energy underdense  ($\delta_{de}<0$) region, while the inflow of quintessential dark energy forms the overdense ($\delta_{de}>0$) one. Going beyond the linear approach (see the next sections) does not change the character of evolution: when inhomogeneities of dark and baryon matter collapse, the density of phantom dark energy continues to decrease in them.

\section{Dark energy in gravitational fields of spherical static objects}

Let us consider now the dynamics of dark energy in the gravitational field of nonrotating spherical static objects. The space-time metric in spherically symmetric coordinates in the general case can be presented as follows:
\begin{equation}
 ds^2 =  e^{\nu(t,r)}d\tau^2 - e^{\lambda(t,r)}dr^2 - r^2\left(d\theta^2+\sin^2\theta d\varphi^2\right).
 \label{s2:sphericsym}
\end{equation}
We assume that metric is completely defined by the spherical static body with radius $R$ and constant density $\overline{\rho}_m$ and pressure $\overline{p}_m\ll \overline{\rho}_m$. In such a case the components of the metric do not depend on time and are determined by Einstein equations with boundary conditions $\lambda|_{r=0}=0$ (see, for example, Ref. \cite{Landau1971}) by:
\begin{equation}
e^{-\lambda(r)} = 1 - \frac{8\pi G}{r}\int\limits_0^{r}\overline{T}^0_0r'^2dr'\,, \quad
\nu(r)+\lambda(r) = 8\pi G\int\limits^r_{\infty}\left(\overline{T}^0_0-\overline{T}^1_1\right)e^{\lambda}r'dr'\,,\label{gcomps:2}
\end{equation}
where $\overline{T}^i_j$ are components of the energy-momentum tensor of the source of the  gravitational field (metric). Since for such a body $\overline{T}^0_0=\overline{\rho}_m$ at $r\le R$ and $\overline{T}^0_0=0$ at $r>R$, outside the object at $r\ge R$ we have $e^{\nu(r)}=e^{-\lambda(r)}=1-r_g/r$, where $r_g=2GM$ is the gravitational radius of object, $M=4\pi\overline{\rho}_m R^3/3$ is its mass. Thus, the metric of 4-space in the neighborhood of the nonrotating spherical static object ($r\ge R$) is the Schwarzschild metric:
\begin{equation}
 ds^2 =  \left(1-r_g/r\right)d\tau^2 - \frac{dr^2}{1-r_g/r} - r^2\left(d\theta^2+\sin^2\theta d\varphi^2\right).
\label{s2:shvartshild}
\end{equation}
For most stars and galaxies $\overline{T}_1^1\ll \overline{T}_0^0$, so we can neglect the component $\overline{T}_1^1=-\overline{p}_m$ in (\ref{gcomps:2}) and obtain the simple space-time metric inside the homogeneous object ($r\le R$),
 \begin{equation}
 ds^2 = \frac{(1-\alpha)^{3/2}}{(1-\alpha r^2/R^2)^{1/2}}d\tau^2 - \frac{dr^2}{1-\alpha r^2/R^2} - r^2\left(d\theta^2+\sin^2\theta d\varphi^2\right),
\label{s2:shvartshild_in}
\end{equation}
where $\alpha\equiv r_g/R$ is constant for given spherical static object. The space-time metric
(\ref{s2:shvartshild}), (\ref{s2:shvartshild_in}) is regular, not singular, in all space-time for objects with $\alpha<1$. Obviously, the top-hat density profile is a very rough approximation for real astrophysical static objects, but it is sufficient for evaluation of the dark energy amount in stars and their vicinities.

Henceforth, we will assume that the space-time metric (\ref{s2:shvartshild}), (\ref{s2:shvartshild_in}) is not affected by the presence of dark energy. The estimations of dark energy density in galaxies and clusters of galaxies  given above provide grounds for such an assumption. Thus, considering the dynamics of dark energy near or inside the spherical static object with metric (\ref{s2:shvartshild}), (\ref{s2:shvartshild_in}), we assume that dark energy is the test component.

The metric (\ref{s2:shvartshild}), (\ref{s2:shvartshild_in}) can be implemented by the observers being at rest with respect to the center of the object with clocks that are synchronized with the clock of the remote observer ($r=r_{\infty}$). The proper time interval of the observer at the distance $r$ from center $d\tilde\tau(r)$ is connected with coordinate interval $d\tau$ as $d\tilde\tau(r)=e^{\nu(r)/2}d\tau$. The proper distance interval along the radius is  $d\tilde{r}=e^{\lambda(r)/2}dr$.

The goal of the analysis of dynamics of the dark energy near or inside the spherical static object with metric (\ref{s2:shvartshild}), (\ref{s2:shvartshild_in}) is to obtain the dependence of its density on the radial coordinate, $\rho_{de}(r)$, and radial component of 3-velocity $v_{de}(r)\equiv d\tilde{r}/d\tilde\tau$ (in units of speed of light), defined as the ratio of the proper space interval to proper time interval at distance $r$ from the center. To obtain the equations of motion for dark energy in such a field, let us represent its 4-velocity $u^i\equiv dx^i/ds$ through 3-velocity $v_{de}(r)$ with:
\begin{equation}
u_{i} = \left\{\frac{e^{\nu(r)/2}}{\sqrt{1-v_{de}^2}},-\frac{v_{de}e^{\lambda(r)/2}}{\sqrt{1-v_{de}^2}},0,0\right\},\quad u^{i}= \left\{\frac{e^{-\nu(r)/2}}{\sqrt{1-v_{de}^2}},\frac{v_{de}e^{-\lambda(r)/2}}{\sqrt{1-v_{de}^2}},0,0\right\},\label{uv}
\end{equation}
so that $u^iu_i=g^{ik}u_iu_k = 1$.
The components of the energy-momentum tensor of dark energy $T_i^k = (\rho_{de}+p_{de})u_iu^k - \delta_i^kp_{de}$ in such a representation are the following:
\begin{equation}
 T_0^0 = \frac{\rho_{de}+v_{de}^2p_{de}}{1-v_{de}^2} = \varepsilon_{de},\quad
 T_0^1 = \frac{\rho_{de}+p_{de}}{1-v_{de}^2}v_{de}e^{\nu(r)/2}e^{-\lambda(r)/2},\quad
 T_1^1 =-\frac{v_{de}^2\rho_{de}+p_{de}}{1-v_{de}^2},\quad
 T_2^2 = T_3^3 = - p_{de}.\label{tei}
\end{equation}
Here, $\varepsilon_{de}$ is the density of dark energy in the observer frame. The equations describing the dynamics of dark energy near the spherical static object are obtained from differential conservation equations in metric (\ref{s2:shvartshild}), (\ref{s2:shvartshild_in}):
\begin{equation}
T^k_{i\,;k}=0. \label{e-c-l}
\end{equation}

Let us consider two important cases: static distribution of dark energy near the objects with different ratios $r_g/R$ and stationary accretion onto the black hole.

\subsection{Static distribution of dark energy in the field of spherical objects}

Here, we will try to answer the question of whether or not the static equilibrium exists in the distribution of dark energy near and inside the nonrotating spherical objects. Static distribution of the dark energy presumes that in each point magnitudes of velocities of the dark energy\footnote{Index $_{de}$ is omitted in the cases in which we have only one component, the dark energy.} $v(r)=0$, and $\rho$ and $p$ are functions of only radial coordinate $r$. The static condition is the equilibrium of the gravitational force and pressure gradient. The equation expressing this condition is the only nontrivial equation in system (\ref{e-c-l}),
\begin{equation}
\frac{dp}{dr}+\frac12(\rho+p)\frac{d\nu}{dr}=0,\label{eq_stat1}
\end{equation}
where $p$ and $\rho$ are unknown functions of $r$ and $\nu$ is given in (\ref{s2:shvartshild}) and (\ref{s2:shvartshild_in}).
For its solutions the model of dark energy must be specified. If it is a simple barotropic fluid, $p(\rho)=w\rho$, $w=const<-1/3$, then the solution of the Eq. (\ref{eq_stat1}) is
\begin{equation}
\rho(r)=\rho_{\infty}\left[e^{\nu(r)}\right]^{-\frac{1+w}{2w}},\label{rho_stat2}
\end{equation}
where $\rho_{\infty}=\rho(r_{\infty})$ is the average dark energy density in а galaxy. The model of dark energy in the form of barotropic fluid, however, is strongly gravitationally unstable because of the negative square of sound speed $c_s^2=w$ [see Eqs. (\ref{delta_de}) and (\ref{V_de}) for a static world and Fig. 7 in Ref. \cite{Novosyadlyj2010} for the FRW background], which contradicts  the observational data on the large scale structure of the Universe.
\begin{figure}
  \begin{center}
    \includegraphics[width=0.49\textwidth]{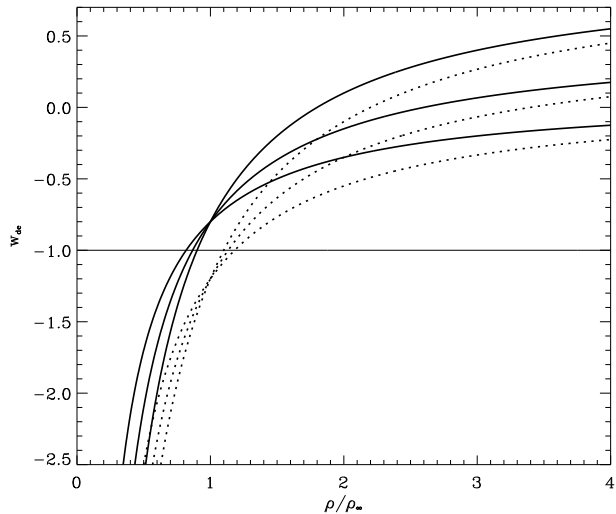}
  \end{center}
  \caption{Dependence of EoS parameter of dark energy $w$ on its density (in units of the background density of dark energy) for models with  $w_{\infty}=-0.8$ (solid lines) and $w_{\infty}=-1.2$ (dotted lines) and the effective sound speed $c_s^2=$1, 0.5, 0.1 (from top to bottom on the right).}
  \label{fig1}
\end{figure}

Scalar fields are prospective models of dark energy (see Refs. \cite{GRG2008,Amendola2010,Wolchin2010,Ruiz2010,Novosyadlyj2013m} and citations therein). Nonzero internal entropy $s$ (entropy density) and the dependence of pressure on density and entropy  $p(\rho,s)$ are inherent for them, which gives the possibility to provide the large negative magnitude of the pressure and positive square of sound speed:
$c_s^2\equiv\delta p/\delta\rho\ge0$. For example, for a field with canonical kinetic term in the Lagrangian ($L=\pm X-U(\phi)$, where
$X\equiv \dot{\phi}^2/2$ is the kinetic term, $U$ is the potential of the field $\phi$), $c_s^2=1$. Among the fields that are used to interpret the observational cosmology data, both the fields varying in time effective sound speed and the fields with constant effective sound speed are considered. Let us consider the easiest case, in which $c_s^2$ does not depend on time and space coordinates. The conditions $\delta p/\delta\rho=c_s^2>0$ and  $p/\rho=w<0$ are both satisfied when the equation $d\rho/\rho=dw/(c_s^2-w)$ is satisfied. It gives simple solution for $w$ for $c_s^2=const$:
\begin{equation}
w=c_s^2-(c_s^2-w_{\infty})\frac{\rho_{\infty}}{\rho}.\label{w-rho}
\end{equation}
Such a model of dark energy has been analyzed also in Refs. \cite{Babichev2004,Babichev2013}. The dependences of EoS parameter (\ref{w-rho}) on its density (in background units), $\rho/\rho_{\infty}$, for models of dark energy with $w_{\infty}=-0.8$ and $w_{\infty}=-1.2$ and three values of square of effective sound speed $c_s^2=$1, 0.5, 0.1 are shown in Fig. \ref{fig1}. As we see, the type and properties of such dark energy can vary in space in correspondence to changes of density: in the regions with density higher than $\rho_{\infty}$, the EoS parameter $w$ becomes $>-1$ even for the model with $w_{\infty}<-1$; in the regions with density lower than $\rho_{\infty}$, the EoS parameter $w$ becomes $<-1$ even for the model with $w_{\infty}>-1$.

Equation (\ref{eq_stat1}) for such a scalar field is easy to integrate and gives the dependence of the density of dark energy on the radial coordinate,
\begin{equation}
\rho(r)=\rho_{\infty}\left(\frac{c_s^2-w_{\infty}}{1+c_s^2}+\frac{1+w_{\infty}}{1+c_s^2}\left[e^{\nu(r)}\right]^{-\frac{1+c_s^2}{2c_s^2}}\right), \label{rho_stat3}
\end{equation}
where $e^{\nu(r)}=1-r_g/r$ outside the spherical object and  $e^{\nu(r)}=(1-\alpha)^{3/2}/(1-\alpha r^2/R^2)^{1/2}$ inside it.

Therefore, the density of dark energy deviates gradually from  $\rho_{\infty}$  when we approach the spherical static object. The character (increasing or decreasing) and  rate of change depend on parameters of the background  dark energy, $c_s^2$ and $w_{\infty}$, as well as on gravitational radius of the central body.
It should be noted that for $c_s^2\rightarrow0$ the solution (\ref{rho_stat3}) loses its meaning. This means that static solutions do not exist for dark energy with $c_s^2=0$: it can only inflow (accrete) on a spherical static object or outflow from its neighborhood. The sign of variation of the density of the dark energy $(\rho(r)-\rho_{\infty})/\rho_{\infty}$ depends on its type: whether it is quintessence or phantom.
\begin{table}
\begin{tabular}{|l|c|c|c|c|c}
  \hline
Object&{Mass $M$ (kg)}&$r_g$(m)&Radius $R$(m)& $(\rho (0)-\rho_{\infty})/\rho_{\infty}$ \\
\hline
Earth-like planet & $6\cdot10^{24}$  & $1\cdot 10^{-2}$ & $6\cdot 10^{6}$  &$\pm2.5 \cdot 10^{-10}$\\
Sun-like star & $2\cdot10^{30}$  & $3\cdot 10^3$ & $7\cdot 10^{8}$  &$\pm6.4 \cdot 10^{-7}$\\
White dwarf & $2\cdot10^{30}$  & $3\cdot 10^3$ & $6\cdot 10^{6}$  &$\pm7.5 \cdot 10^{-5}$\\
Neutron star & $4\cdot10^{30}$ & $6\cdot 10^3$ &$1\cdot 10^{4}$&$\pm 0.3$\\
Galaxy& $3 \cdot 10^{42}$ & $5\cdot 10^{15}$ & $6 \cdot 10^{20}$ & $\pm1.3 \cdot 10^{-6}$\\
Cluster of galaxies & $2\cdot10^{44}$ & $3 \cdot 10^{17}$ & $5\cdot 10^{22}$ & $\pm9 \cdot 10^{-7}$\\
  \hline
\end{tabular}
\caption{Variations of the densities of quintessence (sign ``+'', $w_{\infty}=-0.8$) and phantom (sign ``--'', $w_{\infty}=-1.2$) dark energy with $c_s^2 = 1$ in the centers of different gravitationally bound objects: $(\rho(0)-\rho_{\infty})/\rho_{\infty}$.}
\label{tab1}
\end{table}

\textit{Quintessence DE} ($w_{\infty}>-1$).  With approaching the spherical static object from $r_{\infty}$, the density of such dark energy grows up from $\rho_{\infty}$ to
\begin{equation}
\rho(R)=\rho_{\infty}\left(\frac{c_s^2-w_{\infty}}{1+c_s^2}+\frac{1+w_{\infty}}{1+c_s^2}
\left[1-\alpha\right]^{-\frac{1+c_s^2}{2c_s^2}}\right), \label{rho_stat6}
\end{equation}
at surface $r=R$. So, for $c_s^2>0$, the magnitude of density of quintessential dark energy at the surface of a static object is larger when $\alpha=r_g/R$ is closer to 1. In the case of $\alpha\ll 1$, the expression (\ref{rho_stat6}) can be approximated as
\begin{equation}
\rho(R)\approx\rho_{\infty}\left(1+\alpha\frac{1+w_{\infty}}{2c_s^2}\right). \label{rho_stat6_appr}
\end{equation}
Since for planets, stars, galaxies, and clusters of galaxies  $\alpha\le10^{-6}$, such growth is vanishingly small for the dark energy with $c_s\sim 1$. With the decrease of the effective sound speed the density of dark energy near these objects increases, but a noticeable increase can be reached only for very small $c_s$ ($\le 10^{-3}$), which contradicts, however, the data of cosmological observations. For the compact relativistic objects, neutron stars, and black holes the density of dark energy in their neighborhood can be significant for ``cosmological'' dark energy too. For the neutron star with $\alpha=0.6$, for example, $\rho(R)/\rho_{\infty}\sim 1.15$ for $c_s^2=1$, $\sim16.3$ for $c_s^2=0.1$, and $\sim 10^{19}$ for $c_s^2=0.01$, if the EoS parameter of the background dark energy is $w=-0.8$. As it follows from (\ref{rho_stat6}), when $\alpha\rightarrow 1$ (collapse to the black hole), $\rho(R)/\rho_{\infty}\rightarrow \infty$ even for the scalar field with $c_s^2\sim 1$.

\begin{figure}
  \begin{center}
  \includegraphics[width=0.49\textwidth]{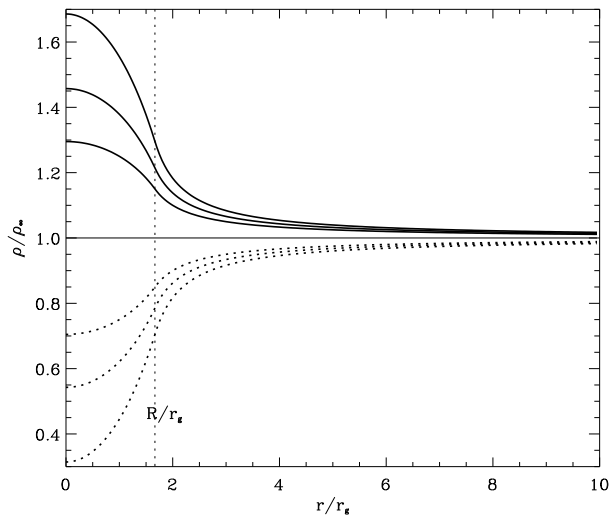}
  \end{center}
  \caption{The dependence of the density of dark energy (in units of the background one) on the distance to center of the homogeneous spherical object (in units of gravitational radius $r_g$) for models of dark energy with $w_{\infty}=-0.8$ (solid lines) and $w_{\infty}=-1.2$ (dotted lines) and three values of the square of the effective sound speed $c_s^2=$1, 2/3, 1/2 (from bottom to top for solid lines and from top to bottom for dotted ones). For the estimation according to (\ref{rho_stat3}) we have taken $R/r_g=5/3$. }
  \label{fig3}
\end{figure}

Inside the spherical static homogeneous object, the density of quintessential dark energy continues to grow according to (\ref{rho_stat3}) and in the center reaches the value
\begin{equation}
\rho(0)=\rho_{\infty}\left(\frac{c_s^2-w_{\infty}}{1+c_s^2}+\frac{1+w_{\infty}}{1+c_s^2}
\left[1-\alpha\right]^{-\frac{3(1+c_s^2)}{4c_s^2}}\right),\label{rho_stat7}
\end{equation}
which is $\approx1.5$ times larger than on the surface (see Fig. \ref{fig3}). In Table \ref{tab1} the estimation of excess of the background density of quintessential dark energy in the centers of different gravitationally bound static objects is given with the assumption that they are spherical, homogeneous, and nonrotating. Obviously, the violation of each of these idealizations will change somehow the dependences of density on the distance to center; however, the order of values and the character of the behaviour will remain the same.

\textit{Phantom DE} ($w_{\infty}<-1$). With approaching the object from distance $r_{\infty}$, the density of phantom dark energy decreases from $\rho_{\infty}$ to (\ref{rho_stat6}) on the surface. For objects with $\alpha\ll 1$, it is lower by the value
$\approx\alpha|1+w_{\infty}|/2c_s^2$ than density at $r_{\infty}$ (Fig. \ref{fig3}). In the center of such objects, these values are yet 1.5 times smaller. In Table \ref{tab1} the estimations of magnitude of $(\rho(0)-\rho_{\infty})/\rho_{\infty}$ for the objects of different classes are presented. For planets, stars, galaxies and clusters of galaxies, such a decrease is very small, $\sim10^{-7}-10^{-10}$,  and  taking into account the smallness of the average dark energy density in the galaxies and clusters of galaxies we can conclude that the influence of  dark energy on the structure of these objects or dynamics of the other bodies in their gravitational field is negligible.   For neutron stars, such a decrease is large. When $\alpha\rightarrow 1$, formally $\rho\rightarrow -\infty$ as it follows from (\ref{rho_stat6}) for $w<-1$. Actually, it may mean that phantom dark energy avoids neutron stars and black holes, $\rho\rightarrow 0$.
Thus, we see that the density of phantom dark energy in the potential wells of the concentration of matter is lower than average and tends to zero when the size of the object approaches the value of gravitational radius. This is the same inherent feature of the phantom dark energy, which we have noted in Sect. II, analyzing the behaviour of perturbations of phantom dark energy in the field of perturbations of dark matter.

It should be also noted that the density changes near and inside the objects, for which estimations in the Table \ref{tab1} are provided, are so small that they do not cause the change of the type of dark energy (from quintessence to phantom or vice versa).

The solutions of the problem of gravitational stability in the gravitational field of the matter inhomogeneity (\ref{delta_de}), (\ref{V_de}) give grounds to state that static solution  (\ref{rho_stat3}) is stable: a small local deviation of density from the equilibrium state will lead to the oscillation around the equilibrium value.

\subsection{Stationary accretion of dark energy onto the black hole}

The density distribution of dark energy near and inside the spherical static objects obtained for the condition of static equilibrium points to the absence of a noticeable influence of the dark energy on the structure and dynamics of motion of bodies in the gravitational fields of planets, stars, galaxies, and clusters of galaxies. However, in the neighborhood of strongly relativistic objects with $R\sim r_g$, the density of the quintessential dark energy can reach large values, at the same time when the density of phantom one ``disappears.'' The possibility of the realization of such condensation/dilution of the quintessence/phantom dark energy in the neighborhood of neutron stars and black holes depends on the rate of its inflow, which we are going to analyze in this subsection.

For the first time the problem of accretion of dark energy onto the black hole with the Schwarzschild metric was solved in Ref. \cite{Babichev2004}, in which the analytical solutions of the problem of dark energy accretion were obtained for the condition that it is a test component, that is, in the absence of its influence on the space-time metric. Nevertheless, the authors have made the conclusion about the decrease of the mass of a black hole as a result of the accretion of phantom dark energy. The results obtained here concerning the behaviour of perturbations of dark energy in the field of matter perturbations (Sec. II) and the distribution of phantom dark energy in the field of static objects (Sec. III.A) point to the fact that a mathematically correct result has been interpreted  incorrectly. To prove this, we carry out the reanalysis of the accretion of dark energy onto the Schwarzschild black hole.

We assume that 1) accretion of the dark energy does not change the space-time metric near the central body, which means that the mass of dark energy that has accumulated in the neighborhood of the central body is negligibly small in comparison to the object mass, and 2) densities of the diffusive components of an interstellar medium (dark and dust matter) are negligibly small near the central body. Such assumptions give reason to solve the problem of stationary accretion of dark energy onto the central body with metric (\ref{s2:shvartshild}).
The solution of such a problem, if it exists, gives the dependence of these quantities on the radial coordinate, on the mass of central body, and on dark energy parameters.

To begin with let us estimate the radius of the region $r_{\infty}$ around the central body, with mass $M$, in which the mass of dark energy $m=4\pi\rho_{\infty}r_{\infty}^3/3\sim M$. For masses $M=10M,\,10^6M$, and $10^9M_{\odot}$ and mean density $\rho_{\infty}=10^{-29}$ g/cm$^3$ the values of such radii are $\sim$250 pc, $\sim$12 kpc, and $\sim$120 kpc, respectively.

The equation of conservation of the energy momentum of dark energy $T^i_{0\,;i}=0$  for (\ref{tei}) in metric (\ref{s2:shvartshild}) gives the differential equation which, is easy to integrate and in the region $r>r_g$ takes the form of
\begin{equation}
r^2\frac{v}{1-v^2}(\rho+p)e^{\nu} = C,\label{ac_eq1}.
\end{equation}
where $C$ is the integration constant determined by the boundary conditions at $r=r_{\infty}$, $\rho(r_{\infty})=\rho_{\infty}$, $w(r_{\infty})=w_{\infty}$, $v(r_{\infty})=v_{\infty}$, and $\nu_{\infty}=\lambda_{\infty}=0$,
\begin{equation}
C=(1+w_{\infty})\rho_{\infty}r_{\infty}^2v_{\infty} \label{C_1},
\end{equation}
Here and below, we assume that $v_{\infty}^2\ll1$. The sign of constant $C$ for a given type of dark energy is determined by the sign of velocity $v_{\infty}$, and for accretion ($v_{\infty}<0$), it is negative for the quintessential dark energy and positive for the phantom one. For blowing out (``wind'' of the dark energy) the signs are opposite. Equation (\ref{ac_eq1}) allows us to evaluate the mass of dark energy that flows inside a sphere of radius $r$ per unit of time. It is defined as $\dot{m}\equiv-4\pi r^2T_0^1$, and following (\ref{tei}), (\ref{ac_eq1}), and (\ref{C_1}), we obtain
\begin{equation}
\dot{m}=-4\pi r_{\infty}^2v_{\infty}(1+w_{\infty})\rho_{\infty}. \label{dot_m_1}
\end{equation}
It is positive for accretion of the quintessential dark energy and negative for accretion of the phantom one. Remember that the solution is obtained for the condition $r_g=const$, which means that the time variation of the mass of central body, $\dot M$, is beyond the scope of this approach.
The correct analysis of the influence of dark energy on the space-time metric was done in Ref. \cite{Dokuchaev2011}.

The second equation can be obtained from another conservation law: $T^i_{1\,;i}=0$. Taking into account (\ref{ac_eq1}) and $u^{\alpha}\ne0$ makes it equivalent to the equation $u^kT^i_{k\,;i}=0$, which implies an equation of motion in the components of the 4-velocity \cite{Babichev2004}:
\begin{equation}
(\rho+p)u^k_{;k}+\rho_{,k}u^k=0. \label{ac_eq2a}
\end{equation}
In the case of stationary accretion described in spherical coordinates, this equation takes the form of
\begin{eqnarray}
\frac{d\rho}{dr}u^1+(\rho+p)\frac{du^1}{dr}+\frac{2}{r}(\rho+p)u^1=0,
\label{ac_eq2}
\end{eqnarray}
where $u^1\equiv dr/ds$.
Its integration gives the equation, which relates the density $\rho$, pressure $p$, and 3-velocity $v$ of the dark energy which stationary inflows in the field of the spherical static object:
\begin{eqnarray}
r^2\frac{ve^{-\lambda/2}}{\sqrt{1-v^2}}\exp{\left[\int_{\rho_{\infty}}^{\rho(r)}\frac{d\rho}{\rho+p(\rho)}\right]}=-A.
\label{ac_eq3}
\end{eqnarray}
The constant $A=-r_{\infty}^2v_{\infty}$ has positive value in the case of accretion, since $v_{\infty}<0$. Equations (\ref{ac_eq1}) and (\ref{ac_eq3}) are equivalent to the Eqs. (1) and (3) from Ref.\footnote{The constants $A$ and $C$ are related with constants in Ref.\cite{Babichev2004} via factor $r_g^2/4$.} \cite{Babichev2004}. For dark energy with equation of state (\ref{w-rho}), the integral in (\ref{ac_eq3}) has the analytical representation, and for a region in which  $v^2\ll1$ the system of equations (\ref{ac_eq1}) and (\ref{ac_eq3}) is easily solved with respect to $\rho$ and $v$:
\begin{equation}
\rho(r)=\rho_{\infty}\left(\frac{c_s^2-w_{\infty}}{1+c_s^2}+\frac{1+w_{\infty}}{1+c_s^2}
\left[1-\frac{r_g}{r}\right]^{-\frac{1+c_s^2}{2c_s^2}}\right),\quad
v(r)=v_{\infty}\left(\frac{r_{\infty}}{r}\right)^2\left(\frac{r}{r-r_g}\right)^{\frac{1+3c_s^2}{2c_s^2}}. \label{rho_ac1}
\end{equation}
The density distribution in this region coincides with the static one, and the amplitudes of density and velocity toward the center increase faster with a smaller value of the effective speed of sound $c_s$. The relativistic regime of accretion ($v\sim 0.1c$) begins far away from object, at $r\sim r_g+3r_{\infty}\sqrt{|v_{\infty}|}$ for $c_s=1$.

From the Eq. (\ref{ac_eq1}) it follows that  during accretion of the dark energy with $c_s=1$ onto the black hole the critical point of the accretion, in which $v=c_s$, is situated on the event horizon, $r_*=r_g$. When $c_s<1$ the critical point is at $r_*>r_g$, but $v\rightarrow 1$ when $r\rightarrow r_g$. The existence of critical point $r_*=(1+3c_s^2)r_g/4c_s^2$ in the accretion flow of dark energy, for which the velocity of flow is equal to effective speed of sound, $v_*\equiv v(r_*)=c_s$, gives us the possibility to estimate the flow of mass $\dot{m}$ as $A=(1+3c_s^2)^{(1+3c_s^2)/2c_s^2}r_g^2/16c_s^3$ \cite{Babichev2004,Babichev2013}. For the model of dark energy with $c_s^2=const>0$ and $w_{\infty}<0$, we obtain
\begin{equation}
\dot{m}=\pi\frac{(1+3c_s^2)^{\frac{1+3c_s^2}{2c_s^2}}}{4c_s^3}(1+w_{\infty})\rho_{\infty}r_g^2.\label{d_m}
\end{equation}
The rate of change of the inflow mass of dark energy is larger for a larger mass of black hole $M$, larger value of background density $\rho_{\infty}$ and smaller value of effective speed of sound $c_s$. For black holes with masses $M=10\,10^6,$ and $10^9 M_{\odot}$ the values of rates of mass change for the dark energy with $c_s=1$ and $w=-0.8$ are equal to $\sim10^{-32},\,\sim10^{-22},$ and $\sim10^{-16}\, M_{\odot}$/year correspondingly. With the decrease of the effective sound speed, the mass inflow of dark energy grows. For example, for $c_s=0.1$ these quantities grow in $\sim10^3$ times. Another characteristic is the time of inflow of the amount of dark energy comparable to the black hole mass $D\tau\approx M/\dot{m}$. In the case of quintessential dark energy, it is  $\sim10^{32},\,\sim10^{22},$ and $\sim10^{16}$ years respectively, which is much more than the age of the Universe. The total ``disappearance'' of phantom dark energy in the sphere with $r=3r_g$ around such black holes or the increase by 
two times
the mass of quintessential dark energy in this volume happens in a very short time, $\Delta\tau\approx12r_g/|1+w_{\infty}|=1.2\cdot10^{-3}M/M_{\odot}$ s, since for mean density $\rho_{\infty}\sim10^{-29}$ g/cm$^3$ this mass is small.

\begin{figure}
  \begin{center}
  \includegraphics[width=0.49\textwidth]{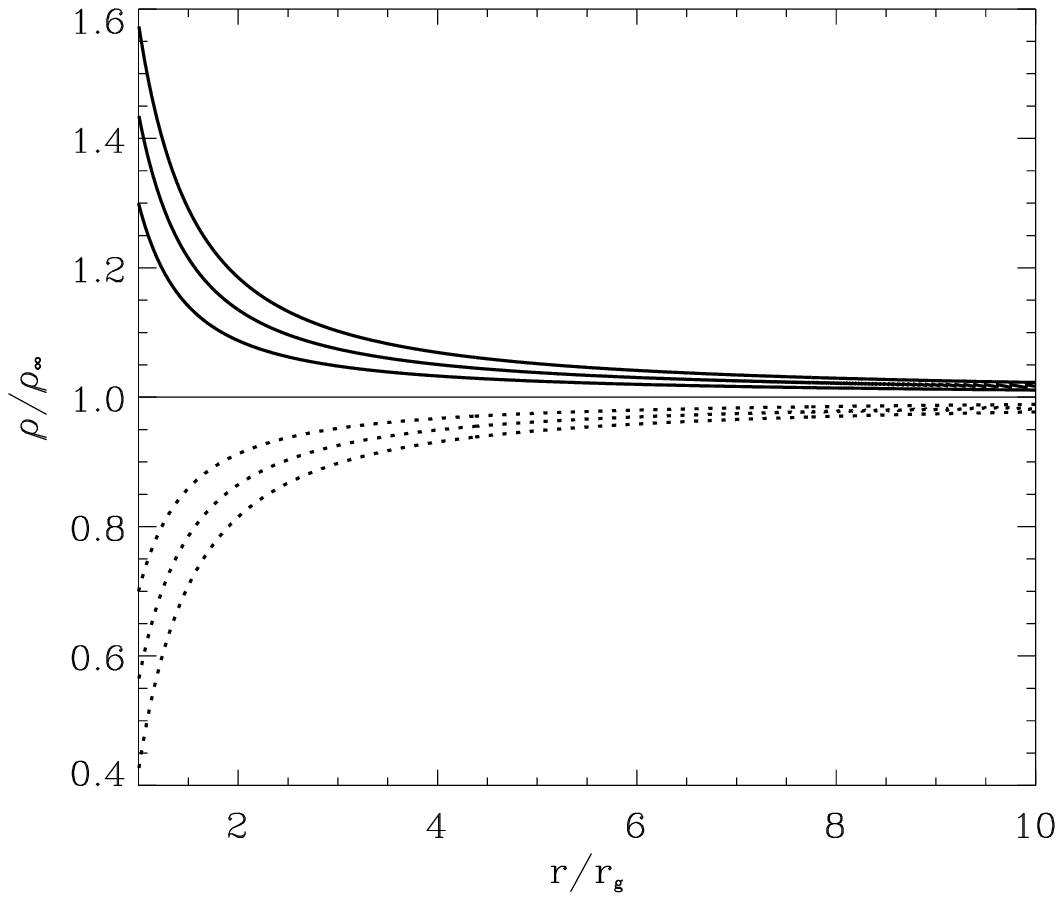}
  \includegraphics[width=0.49\textwidth]{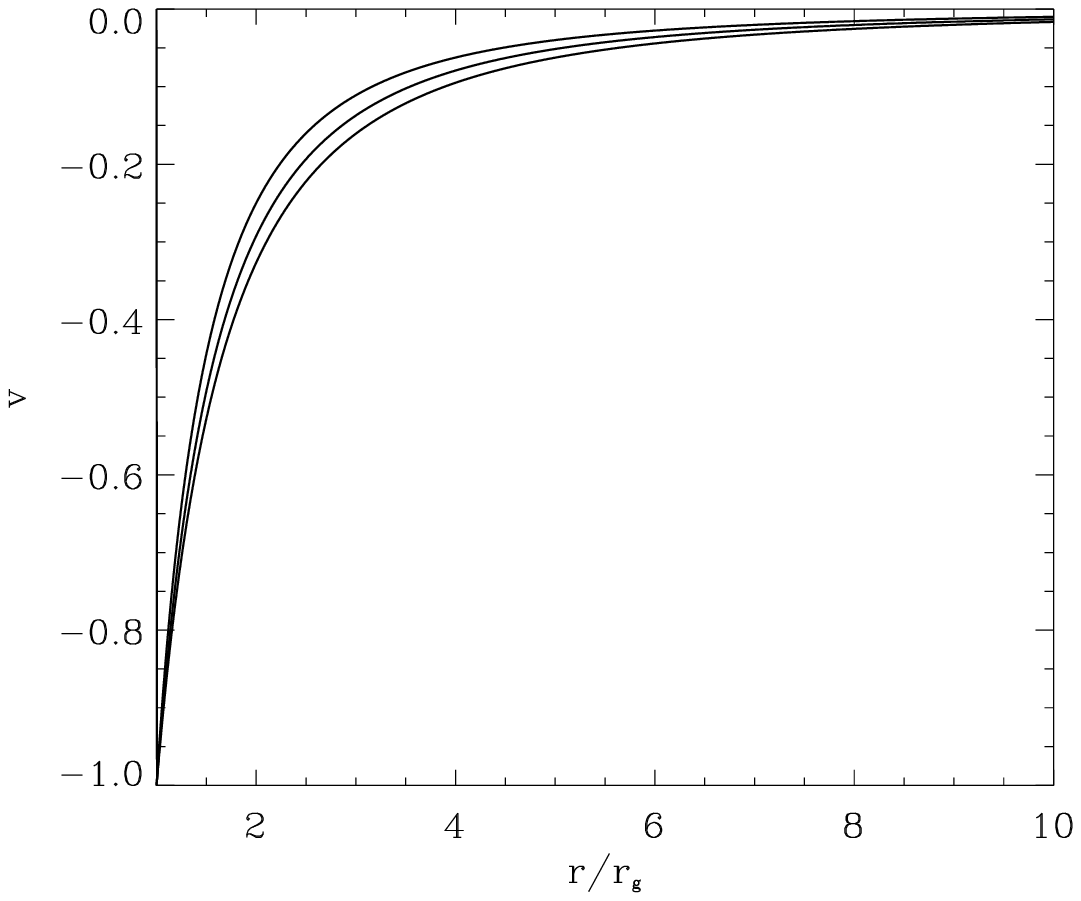}
  \end{center}
  \caption{The dependences of dark energy density (in units of $\rho_{\infty}$, left) and radial component of the 3-velocity (in units of speed of light, right) on the distance to the center of black hole (in units of gravitational radius $r_g$) for the models of dark energy with $w_{\infty}=-0.8$ (solid lines) and $w_{\infty}=-1.2$ (dotted lines) and three values of the square of the effective sound speed $c_s^2=$1, 2/3, 1/2 (from bottom to top in the left panel for quintessential dark energy and from top to bottom for a phantom; in the right panel from top to bottom).}
  \label{fig4}
\end{figure}

Let us compare the density distributions of dark energy near the relativistic objects in the case of static equilibrium  (\ref{rho_stat3})  and in the case of stationary accretion. In the latter case it can be obtained by solving the Eqs. (\ref{ac_eq1}) and (\ref{ac_eq3}), which for dark energy with  $c_s^2=const>0$ and $w_{\infty}<0$ give the algebraic system of equations for $\rho(r)$ and $v(r)$:
\begin{eqnarray}
\frac{v}{1-v^2}\left[\frac{(1+c_s^2)\rho-(c_s^2-w_{\infty})\rho_{\infty}}{(1+w_{\infty})\rho_{\infty}}\right]=-\frac{(1+3c_s^2)^{\frac{1+3c_s^2}{2c_s^2}}}{16c_s^3}\frac{r_g^2}{r^2}e^{-\nu},\label{ac_1}\\
\frac{v}{\sqrt{1-v^2}}\left[\frac{(1+c_s^2)\rho-(c_s^2-w_{\infty})\rho_{\infty}}{(1+w_{\infty})\rho_{\infty}}\right]^{\frac{1}{1+c_s^2}}=
-\frac{(1+3c_s^2)^{\frac{1+3c_s^2}{2c_s^2}}}{16c_s^3}\frac{r_g^2}{r^2}e^{\lambda/2}. \label{ac_2}
\end{eqnarray}
It has solutions in analytical functions for $c_s^2$=1/4, 1/3, 1/2, 2/3, 1, 3/2 and 2 \cite{Babichev2013}.
Indeed, in the case of $c_s^2=1$ the system (\ref{ac_1}) and (\ref{ac_2}) has two solutions, and the one that describes the stationary accretion of dark energy is as follows:
\begin{eqnarray}
v(r)&=&-\frac{r_g^2}{r^2}, \label{ac_v_1}\\
\rho(r)&=&\rho_{\infty}\left[\frac{1-w_{\infty}}{2}+\frac{1+w_{\infty}}{2}\left(1+\frac{r_g}{r}\right)\left(1+\frac{r_g^2}{r^2}\right)\right].\label{ac_rho_1}
\end{eqnarray}
The critical point of accretion for such dark energy is at the event horizon ($r_*=r_g$). One can note that $v_{\infty}=-r_g^2/r_{\infty}^2$ and according to (\ref{C_1}) $C=-(1+w_{\infty})\rho_{\infty}r_g^2$ and according to (\ref{ac_eq3}) $A=r_g^2$.

In the case of $c_s^2=2/3$, the system of equations (\ref{ac_1}) and (\ref{ac_2}) has three solutions, the next two of which describe the stationary accretion,
\begin{eqnarray}
v(r)&=&-\sqrt{\frac{2}{3}\cos{\left(\frac{1}{3}\arccos\left(1-\frac{27}{2}\kappa^4\frac{r_g^8}{r^8}\left(1-\frac{r_g}{r}\right)^{-1}\right)-k\frac{2\pi}{3}\right)}+\frac{1}{3} }, \label{ac_v_0.67}\\
\rho(r)&=&\rho_{\infty}\left[\frac{2-3w_{\infty}}{5}-3\frac{1+w_{\infty}}{5}\kappa\left(1-\frac{r_g}{r}\right)^{-1}\frac{r_g^2}{r^2}\frac{1-v^2(r)}{v(r)}\right], \label{ac_rho_0.67}
\end{eqnarray}
where $\kappa\equiv(1+3c_s^2)^{(1+3c_s^2)/2c_s^2}/16c_s^3\approx1.36$, $k=0$ for $r_g\le r\le r_*$, $k=1$ for $r>r_*$, and $r_*=1.125r_g$.

In the case of $c_s^2=1/2$, the system of equations (\ref{ac_1}) and (\ref{ac_2}) has four solutions, the following two of which describe the stationary accretion:
\begin{eqnarray}
v(r)&=&-\sqrt{\frac{1}{2}\pm\sqrt{\frac{1}{4}-\left(\frac{5}{4}\right)^5\left(1-\frac{r_g}{r}\right)\frac{r_g^4}{r^4}}}, \label{ac_v_0.5}\\
\rho(r)&=&\rho_{\infty}\left[\frac{1-2w_{\infty}}{3}-2\frac{1+w_{\infty}}{3}\left(1-\frac{r_g}{r}\right)^{-1}\frac{r_g^2}{r^2}\frac{1-v^2(r)}{v(r)}\right] \label{ac_rho_0.5}
\end{eqnarray}
(sign ``$+$'' for $r_g\le r\le r_*$, sign ``$-$'' for $r>r_*$; $r_*=1.25r_g$). The distributions of density and velocity of dark energy in the  neighborhood of a black hole in the case of stationary accretion and different values of parameters of dark energy  are shown in Fig. \ref{fig4}. As we see, the density distributions of dark energy at $r\ge 1.5r_g$ are close to those in the static distribution. It is interesting that velocity $v$ is different for different values of the effective sound speed $c_s$ but does not depend on $w_{\infty}$.

As in the case of static density distribution near the spherical objects (\ref{rho_stat3}), the density of quintessential dark energy near the black hole is larger for a smaller effective sound speed. The existence of the event horizon of a black hole does not allow the static solutions for dark energy in its neighborhood, because, according to the solution, for the static distribution (\ref{rho_stat3}) for quintessential dark energy, $\rho\rightarrow\infty$, and for phantom, formally $\rho\rightarrow-\infty$ when $R\rightarrow r_g$, while according to the solutions (\ref{ac_rho_1}), (\ref{ac_rho_0.67}), (\ref{ac_rho_0.5}) at the event horizon the dark energy density has a finite value.
Nevertheless, the estimation of characteristic time $\Delta\tau$ of the dark energy mass change in the sphere with radius $3r_g$ as a result of the stationary accretion shows that establishing the static equilibrium distribution of the dark energy density around spherical compact object (\ref{rho_stat3})-(\ref{rho_stat7}) is possible.
\begin{figure}
  \begin{center}
  \includegraphics[width=0.49\textwidth]{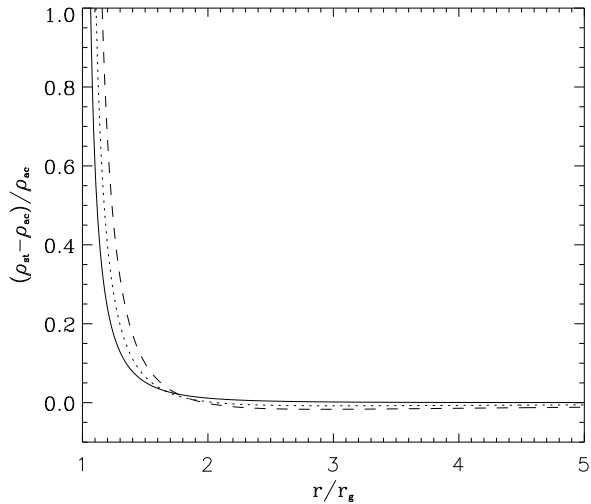}
  \end{center}
  \caption{The relative differences of densities  $(\rho_{st}-\rho_{ac})/\rho_{ac}$ of quintessential dark energy ($w_{\infty}=-0.8$) with statical distribution $\rho_{st}(r)$ (\ref{rho_stat3}) near the relativistic object $R=r_g+\epsilon$ and stationary accretion onto the black hole $\rho_{ac}(r)$ (\ref{ac_rho_1}), (\ref{ac_rho_0.67}), (\ref{ac_rho_0.5}) for models with $c_s^2=$1 (solid line), 2/3 (dotted lines), and 1/2 (dashed line). For phantom dark energy, the relative differences of densities have close magnitudes but the opposite sign.}
  \label{fig5}
\end{figure}

In Fig. \ref{fig5}, the relative differences of densities of quintessential dark energy $\rho_{st}(r)$ for static distribution (\ref{rho_stat3}) near the relativistic object with  $R=r_g+\epsilon$ and distributions for stationary accretion onto a black hole $\rho_{ac}(r)$ (\ref{ac_rho_1}), (\ref{ac_rho_0.67}), (\ref{ac_rho_0.5}) are given for the models with $c_s^2=$1, 2/3, and 1/2, correspondingly. At $r>1.5r_g$ for such values of the squared effective sound speed, $\rho_{st}(r)$, and $\rho_{ac}(r)$ practically coincide, but when $r\rightarrow r_g$, the ratio $\rho_{st}/\rho_{ac}\rightarrow\infty$. For phantom dark energy, the relative differences of densities have close absolute magnitudes but the opposite sign.

\section{Discussion}

{\it Dark energy in galaxies and clusters of galaxies.} In the analysis of dynamics of dark energy in the static fields of gravitationally bound systems we have used the model of dark energy with EoS parameter (\ref{w-rho}), which admits any $c_s^2>0$ and any $w<-1/3$.
In the static world of galaxies or clusters of galaxies the average density of dark energy does not change with time, unlike the dark energy outside them where it takes part in the Hubble expansion; however, it changes in space in accordance with change of the gravitational potential caused by matter inhomogeneities. Since the dark energy does not take part in the process of virialization and establishing of the dynamical equilibrium in gravitationally bound systems, it is natural to assume that the value of density of dark energy in galaxies or clusters of galaxies is the same  as it was at cosmological background at the moment of their formation, and this explains the relation $\overline{\rho}_{de}^{(gal,\,cl)}\ll \overline{\rho}_{m}^{(gal,\,cl)}$. Together with relation $\delta_{de}\ll\delta_m$, it justifies the approximation for the testing of dark energy in the analysis presented here.

The assumption about static dark energy or constant dark energy density at infinity   ($\rho_{\infty}=const$) in the static world is quite reasonable. Indeed, subdominant dark energy in the space-time with constant 4-curvature (gravitation Lagrangian) without special assumptions about its time dependence will be stationary since it is gravitationally stable.
In the case of spherical symmetry, the Birkhoff theorem can be applied to support such an assumption for gravitationally bound systems in the cosmological background. We can suppose also that $\rho_{\infty}(t)$ is related with dark energy density at the cosmological background. But characteristic time $t_{c}=(\dot{\rho}_{\infty}/\rho_{\infty})^{-1}=|3(1+w)H_0|^{-1}$ is essentially larger than characteristic times for galaxy objects and does not change essentially the results obtained here. So, we can call this approximation as quasistatic.

{\it Scalar field model of dark energy.} The example of dark energy model with $c_s^2=const>0$ and $w<0$ is the scalar field $\phi$ with Lagrangian \cite{Fang2007} $L=\pm X^{\beta}-U(\phi)$, where $X\equiv \dot{\phi}^2/2$ is the kinetic term and $U$ is the potential, both are functions only of space coordinates (${\beta}\equiv(1+c_s^2)/(2c_s^2)$, sign ``$+$'' is for the quintessential field, and sign ``$-$'' is for the phantom one). The dynamics of the scalar field with a similar Lagrangian in the expanding Universe was analyzed in Ref. \cite{Sergijenko2014}.
In the static world of gravitationally bound objects, the field variable $\phi$, kinetic term $X$, and potential $U$ can be represented via phenomenological variables $\rho$, $w$, and $c_s^2$ as:
\begin{equation}
\phi=\pm\sqrt{2X}\tau+C, \quad X=\left(\pm c_s^2\rho\mp c_s^2\frac{c_s^2-w_{\infty}}{1+c_s^2}\rho_{\infty}\right)^{\frac{2c_s^2}{1+c_s^2}}, \quad
U=\frac{(1-c_s^2)\rho+(c_s^2-w_{\infty})\rho_{\infty}}{2}, \nonumber
\end{equation}
where $C$ is an arbitrary constant. Let us notice that the field variable $\phi$ can play the role of time. The interesting fact is that for the scalar field with $c_s^2=1$ the potential has constant value in the whole volume of the galaxy or cluster of galaxies,
$U=(c_s^2-w_{\infty})\rho_{\infty}/2$. Here, $\rho_{\infty}$ and $w_{\infty}$ are the density and EoS parameter at the distance $r_{\infty}$, where they are treated as average for the galaxy and related with cosmological dark energy at the epoch of structure formation.

{\it Space variation of dark energy density in galaxies.}  In galaxies and clusters of galaxies the scalar field with parameters analyzed here is gravitationally stable outside the matter inhomogeneities. There, its density and velocity can only oscillate with small amplitude of deviation from their mean values. The dark energy with $c_s^2\sim1$ is also gravitationally stable at cosmological background (see Ref. \cite{Novosyadlyj2010} and citations in it). In the gravitational field of the matter overdensities, it inflows at the linear stage of their evolution (Sec. II), forms the static distribution of density around static objects  (Sec. III.A), or inflows onto black holes (Sec. III.B). The rate of the accretion of dark energy in a Schwarzschild field is too small to have the observational manifestations inside or near the clusters of galaxies, galaxies, and stars, since the values of dark energy density in their vicinities differ only slightly from the background one, which is small. However, the
dependence of the density of dark energy in vicinities of spherical objects on its effective sound speed is important: the density is higher for lower effective sound speed (Fig. \ref{fig3} and \ref{fig4}). Qualitatively, these results are quite general for the dynamical quintessential or phantom dark energy, but quantitatively, they depend on the values of parameters of a specific scalar field model.

{\it Phantom dark energy: Accretion or wind?} The results of the analysis of phantom dark energy accretion onto the black hole, which are obtained on the base of differential equations for energy-momentum conservation, are complicated for interpretation in the framework of usual imagination: when $v(r)<0$, which means the inflow onto the center, $\dot{m}<0$, which means a decrease of the amount of dark energy. We do not know by which physical method the rate or direction of the dark energy flow can be measured; however, the method of the measurement of change of the mass in a sphere of certain radius can be proposed easily. That is why the term ``accretion'' is better to define as $\dot{m}>0$. Thus, the matter condensations ``crowd out'' the phantom dark energy from the space volume where they are located.  The physical reason for the phenomenon is the negative inertial mass of phantom dark energy per unit volume, $\rho_{de}^{inertial}=\rho_{de}+p_{de}=(1+w_{de})\rho_{de}<0$, this follows, for example, from 
Eqs. (\ref{eq_stat1}), (\ref{ac_eq2a}), or (\ref{ac_eq2}).

\textit{Application of the results.} Current cosmological data give no possibility to constrain the value of the effective sound speed \cite{Sergijenko2014}; the posterior function is flat for the top-hat prior [0,1]. But here we see that the very small value of $c_s^2$ must be excluded since the large value of dark energy density would influence the potential wells of stars and significantly change the parameters of Keplerian orbits of planets or components of multiple systems. The absence of such a influence in the range of measurement  accuracy for planets of the Solar System, for example, should give the lower limit for the effective sound speed of dark energy of the analyzed type.
Therefore, the search for observable manifestations of dark energy near the relativistic objects, neutron stars and black holes looks like a the promising task.

\section{Conclusions}

The analysis of dynamics of dark energy in the static field of a gravitationally bound systems has shown that density of dark energy in the galaxy clusters, in the galaxies, and in the vicinity of and inside the stars is by few orders smaller than the mean density of dark matter. It has shown that quintessential ($-1<w<-1/3$) and phantom ($w<-1$) dark energy in the static world of galaxies and clusters of galaxies is gravitationally stable: under the influence of self-gravitation, it can only oscillate.  In the gravitational fields of dark matter perturbations, it can inflow monotonically, but the amplitudes of perturbations of density and velocity remain small at all scales interesting for astrophysics. It has also been shown that the accretion of phantom dark energy in the region of overdensities of dark matter causes the formation of the negative perturbation of the density of dark energy.

The behaviour of dark energy in the gravitational field of static nonrotating spherical objects has been studied. The static distribution of dark energy density in their vicinity and inside them has been obtained [Eqs. (\ref{rho_stat2})-(\ref{rho_stat7})]. It has been found that the dark energy density differs very slightly from the mean one; the magnitude of relative deviation of the density, $\delta_{de}=\rho_{de}(0)/\overline{\rho}_{de}-1$, has maximum in the center and depends on the EoS parameter $w_{\infty}$, and the effective sound speed $c_s$ of dark energy as well as on the gravitational radius of central body $r_g$ and its relative size $\alpha^{-1}=R/r_g$. Its sign is positive for the quintessential scalar field and negative for the phantom one. For stars similar to the Sun,  $\delta_{de}\sim10^{-7}$, and for neutron stars, $\sim10^{-1}$. When $c_s\sim1$ and $w\sim-1\pm0.1$, the density inside and near the compact object is small and barely differs from the average value. The difference becomes 
significant when $c_s\rightarrow 0$ or $r\rightarrow r_g$.

The solutions of equations describing the stationary accretion of dark energy as a test component onto the Schwarzschild black hole have been obtained
[Eqs. (\ref{rho_ac1})-(\ref{ac_rho_0.5})] and analyzed.  It was shown that the mass of dark energy, which crosses the sphere of radius $r>r_g$ toward the center, is determined by the parameters of dark energy and the square of the black hole event horizon. The rate of change of mass is a positive quantity in the case of quintessential dark energy and negative in the case of the phantom one.

\acknowledgments
This work was supported by the project of Ministry of Education and Science of Ukraine (state registration number
0113U003059) and research program ``Scientific cosmic research'' of the National Academy of Sciences of Ukraine (state registration number 0113U002301).

\end{document}